\documentstyle[twocolumn,aps]{revtex}

\begin{document}
\draft
\title{ Irrversibility of entanglement
manipulations: Vagueness of the entanglement of cost and
entanglement of distillation} 
\author{ Won-Young Hwang \cite{email} and
 Keiji Matsumoto \cite{byline}}
\address{IMAI Quantum Computation and Information Project,
ERATO, Japan Science and Technology Corporation,
Daini Hongo White Building 201, 5-28-3, Hongo, Bunkyo,
Tokyo 133-0033, Japan}
\maketitle
\begin{abstract}
We show that the entanglement of cost  
and entanglement of distillation can be vague when
we consider
a more general form of entanglement manipulation in which
we collectively deal with not only states of our concern but also
other states.
We introduce the most general entanglement manipulation
in which the formation and distillation
can be simultaneously performed.
We show that in a certain case entanglement manipulations are
reversible with respect to the most general entanglement 
manipulation. This broadens our scope of vision on
 the irreversibility of entanglement manipulations.

\end{abstract}
\pacs{03.67.-a}
\narrowtext
Entanglement is a key ingrediant in quantum information 
processing. Without full entanglement, 
we cannot expect the 
exponential speedup in quantum computation over the classical 
one in the case of pure states \cite{jozs}. 
In quantum communications like superdense coding \cite{ben3} 
and quantum teleportation \cite{ben4}, entanglement is
indispensible.
At the same time the 
entanglement is a valuable resource that should not be wasted,
since entanglement can only be obtained by (costly) non-local 
operations. 
For the better understanding and manipulation of the entangled 
states, we should classify them as well as 
possible. Quantification of the entanglement, 
namely the measure of entanglement is, therefore, a central issue
 in quantum information theory \cite{ben2,vede}. 

Entanglement of cost $E_C$ and entanglement of distillation $E_D$ 
 were proposed by Bennett {\it et al.} \cite{ben2}. These are
 an important pair of measures because they are intuivitely motivated.
On the other hand, irreversibility in entanglememt manipulations
is another important issue. The question is that 'Can we distill
as much entanglement from a state $\rho$ as we have 
consumed for the 
formation of it, with only quantum local operations with classical 
communications (LOCCs)?'. 
If this is the case,
we can reverse the formation process by the
distillation process without loss of entanglement. Then
we can say that the entanglement manipulation is reversible. 
In finite-copies case, entanglement manipulation
is shown to be irreversible both
for pure \cite{lo,niel,vid3} and mixed states \cite{vede}.
However, in asymptotic case where infinitely many identical
 copies are dealt with,  entanglement manipulation of
 the pure states is shown to be reversible,
 namely we have $E_C(\rho)= E_D(\rho)$ for
 pure states \cite{bene}. However, it has been
 conjectured that the entanglement manipulation is irreversible
 for mixed states even in the asymptotic case.
 Indeed it has been
 shown to be case in a few examples \cite{vida,vid2,mats}.
It is often
 said that the irreversiblity in the asymptotic manipulation
 implies a genuine irreversibility
 \cite{horo}. 
Discovery of a new genuine irreversiblity has a large implication on
 natural sciences. 
However, the irreversibility depends on how general the
entanglement manipulation that we consider is.
In the asymptotic manipulation, we only deal with infinitely many 
copies of a state that we are concerned about. That is, we do not
consider other states.
However,
due to (possible) non-additivity of entanglement,  we need to
consider a more general form of entanglement manipulation in which
we collectively deal with both the states of our concern and
other states.
We will see how the generalization broadens our scope of vision
 on irreversibility of mixed state entanglement
 manipulation \cite{vida,vid2,mats}. 
For example, we find that it is 
reversible with respect to a general entanglement manipulation
in a certain case. 

This paper is organized as follows.
First, we introduce the entanglement of cost, entanglement of
distillation, and the irreversibility.
Next, we show that we should consider
 collective-formation and collective-distillation
 in the case of mixed state.
With respect to these, the 
entanglement of cost and entanglement of distillation can be vague.
Next, we consider a more general entanglement 
manipulation, that of `the center of
entanglement'. We discuss the irreversibility with respect to 
the center of entanglement; The formation and 
distillation processes are reversible with respect to
the center of entanglement in a certain case. 

The measure of entanglement for the bipartite pure state 
$|\psi \rangle \langle \psi|$ is given by
\begin{equation}
\label{a}
E(|\psi \rangle \langle \psi |)= 
S(\mbox{Tr}_B |\psi\rangle \langle \psi| ),
\end{equation}
where $S(\rho)$ is the von Neumann entropy of $\rho$, i.e., 
$S(\rho)= -\mbox{Tr}(\rho \log_2 \rho)$ and $B$ denotes the 
second party of two parties Alice and Bob \cite{bene}.

Entanglement of cost $E_C$ of a state $\rho$ is
 inf.
of $N/M$ in the limit of large $N$, when we create the state by an
 (approximate) transformation like 
\begin{equation}
\label{b}
|\Psi^-\rangle \langle \Psi^-|^{\otimes N} \rightarrow
\rho^{\otimes M}
\end{equation}
with LOCCs. 
Here $N,M$ are integers and $|\Psi^-\rangle= (1/\sqrt{2})
(|0\rangle_A |1\rangle_B-|1\rangle_A |0\rangle_B)$
is one of the Bell states. Entanglement of distillation $E_D$ of
 a state $\rho$ is 
 sup. of $N^{\prime}/M$ in the limit 
of large $N^{\prime}$, when we distill the state $\rho$ by an
 (approximate) transformation like 
\begin{equation}
\label{c}
\rho^{\otimes M} \rightarrow |\Psi^-\rangle \langle
 \Psi^-|^{\otimes N^{\prime}}
\end{equation}
with LOCCs. (For more rigorous definitions see Refs. \cite{vida,hayd}.) 

The irreversibility problem is whether it can be $N=N^{\prime}$
 in a manipulation 
\begin{equation}
\label{d}
|\Psi^-\rangle \langle \Psi^-|^{\otimes N} \rightarrow
 \rho^{\otimes M}\rightarrow |\Psi^-\rangle \langle 
\Psi^-|^{\otimes N^{\prime}}.
\end{equation}
In the case of finite numbers ($N,N^{\prime}$, and $M$) of copies,
 it is irreversible or $N > N^{\prime}$ \cite{vede}.  
In the asymptotic case where infinitely many copies are considered,
 however, due to
(possible) non-additivity of entanglement, it might be reversible.
 Recently, Vidal and Cirac \cite{vida,vid2} and 
Matsumoto {\it et al.} \cite{mats} have shown that
 the asymptotic entanglement manipulation is irreversible,
 that is, $E_C(\rho) > E_D(\rho)$, for certain states $\rho$'s.
However, the irreversibility of other general states
 is still an open problem.

However, it is due to collective-manipulation that entanglement 
manipulation with finite copies can be different
 from that with infinite copies, the asymptotic case.
 Collective-manipulation is
 considered in the asymptotic manipulation of the pure 
 states \cite{bene,pope}: Entanglement manipulation of a single
 copy of
 pure states is highly irreversible \cite{lo,niel,vid3}. That is, except
 for local unitary operations, we lose certain amount of
 entanglement during the
 manipulation. However, if we collectively manipulate infinitely many 
 copies of pure states, the amount of the lost entanglement is
 negligible.
In other words, {\it the more general manipulation we
 use, the less entanglement we lose}.

In the same way,  
collective-formation like 
\begin{equation}
\label{h}
|\Psi^-\rangle \langle \Psi^-|^{\otimes N}
 \rightarrow \rho_1^{\otimes M_1}\otimes \rho_2^{\otimes M_2}
\otimes \cdot \cdot \cdot
\end{equation}
and collective-distillation like
\begin{equation}
\label{i}
\rho_1^{\otimes M_1}\otimes \rho_2^{\otimes M_2}\otimes
 \cdot \cdot \cdot \rightarrow  |\Psi^-\rangle \langle
 \Psi^-|^{\otimes N^{\prime}}.
\end{equation}
can be more advantageous than entanglement manipulation where 
each state $\rho_i$ ($i=1,2,3,...$) is separately formated and distilled,
respectively.
For example, in a
 collective-formation 
 $|\Psi^-\rangle \langle \Psi^-|^{\otimes N} \rightarrow
 \rho_1^{\otimes M_1}\otimes \rho_2^{\otimes M_2}$, it can be 
that $N< M_1 E_C(\rho_1)+M_2 E_C(\rho_2)$.
Here $E_C$ is
 the entanglement of cost that is defined with respect to
 separate entanglement manipulations 
$|\Psi^-\rangle \langle \Psi^-|^{\otimes n} \rightarrow
\rho_j^{\otimes m}$ ($j=1,2$).
However, in these cases we can see 
that {\it there is no unique way of assigning 
entanglement of cost or entanglement of distillation
to each state $\rho_i$}.
In other words, entanglement of creation and entanglement of
distillation are vague, respectively,
when we deal with collective-formation and collective-distillation 
 of many copies of states that include not only a state whose
entanglement of cost and entangement of distillation 
 we want to define but also other different states.
In collective manipulation where we deal with many copies of a state
whose entanglement of cost and entanglement distillation 
we want to define, there is no such vagueness problem.
There is a unique solution, that is, to equally assign 
entanglement to each state \cite{bene}.
A search for a solution for how to distribute the
entanglement of cost and entanglement of distillation
to different states is worthwhile.
It can be that a solution will be given by a certain kind of
self-consistency condition.
The irreversibility problem now is whether we can
close the loop of the transformations in Eq. (\ref{h}) and 
(\ref{i}), that is, whether $N=N^{\prime}$.
This problem can be different from that of 
$|\Psi^-\rangle \langle \Psi^-|^{\otimes N} \rightarrow
 \rho_i^{\otimes M}\rightarrow |\Psi^-\rangle \langle 
\Psi^-|^{\otimes N^{\prime}}$.

Now let us consider the most general entanglement manipulation,
 namely that of  `banks of entanglement'. 
Let us assume that a pair of banks of entanglement were 
established on the Earth and Mars. A pair 
of customers, say E$_1$ and M$_1$, who are respectively
 on the Earth and Mars,
 will make an order for banks to do certain desired manipulations
 on their pairs of qubits at each bank. 
Using the banks is advantageous because
collective manipulations can  possibly save entanglement,
 as we have discussed. 
The more orders the banks gather before they perform a big
collective manipulation, the more entanglement they can save. 
Moreover, the banks can 
adopt all possible form of collective manipulations like the
 catalytic manipulation \cite{jona}, Morikoshi's \cite{mori},
 catalytic manipulation with unlimited amount of  bound
 entanglement \cite{rain,rai2,ved2}, and the asymptotic
 generalization of catalytic manipulation \cite{ved3}.
In fact, concept of the banks has been implicit in previous
discussions.  But there was a restriction
 that the formation and distillation are separately
 performed.
However, the banks of entanglement do not have to keep the
restriction.
More general transformation that the banks perform is 
\begin{eqnarray}
\label{j}
|\Psi^-\rangle \langle \Psi^-|^{\otimes N}&& \otimes
\rho_1^{\otimes M_1}\otimes \rho_2^{\otimes M_2}\otimes
 \cdot \cdot \cdot \nonumber\\
\rightarrow &&
{\rho_1^{\prime}}^{\otimes M_1}\otimes {\rho_2^{\prime}}^{\otimes 
M_2}\otimes \cdot \cdot \cdot \otimes |\Psi^-\rangle \langle
\Psi^-|^{\otimes N^{\prime}},
\end{eqnarray}
where formation and distillation are performed simultaneously.

Now let us consider the irreversibility problem. 
A certain customer will give the banks an order for 
formation of a state $\rho_1$ from the Bell states. 
The other will give the banks an order for 
distillation of the Bell state from another state $\rho_2$. 
Assuming a lot of customers who have different orders,
we can introduce distribution functions $P_i^f$ and $P_i^d$
for each state $\rho_i$ ($i=1,2,3, ...)$.
Here $P_i^f$ and $P_i^d$ are 
 frequencies that formation and distillation of a
state $\rho_i$ is ordered, respectively.
Let us consider a case in which $P_i^f=P_i^d$. Here
 what the banks have to do is only to appropriately
re-distribute the qubits to the customers. Thus it is obviously 
reversible. It is not that we can always get the best case.
However, even when the distribution functions are not the same but 
only partially overlap,
the amount of lost entanglement will be decreased.

In conclusion,
the irreversibility in entanglement manipulations is dependent
upon how general manipulation we deal with. 
We need to
consider a more general form of entanglement manipulation in which
we collectively deal with not only states of our concern but also
other states.
 With respect to the general manipulation, the 
entanglement of cost $E_C$ and distillation $E_D$ can be vague.
We introduced the most
 general entanglement manipulation, namely that of the banks of
 entanglement, in which the formation and distillation
 do not have to be separatedly performed. 
We showed that in a certain case entanglement manipulations are
reversible with respect to the most general entanglement 
manipulation. This broadens our scope of vision on
the irreversibility of entanglement manipulations.

\acknowledgments
We are very grateful to Prof. Hiroshi Imai and
Japan Science and Technology Corporation for financial supports. 
We are also very grateful to Dr. G. Vidal for discussions.

\end{document}